\documentclass[12pt]{iopart}

\usepackage[dvipdfm]{graphicx}
\begin{document}

\title{Carrier-phase Two-Way Satellite Frequency Transfer over a Very Long Baseline}

\author{M Fujieda$^1$, D Piester$^2$, T Gotoh$^1$, J Becker$^2$, M Aida$^1$ and A Bauch$^2$}

\address{$^1$National Institute of Information and Communications
Technology, Koganei, Japan $^2$Physikalisch-Technische Bundesanstalt (PTB),
Bundesallee 100, 38116 Braunschweig, Germany}
\ead{miho@nict.go.jp}
\begin{abstract}

\end{abstract}
In this paper 
we report that carrier-phase two-way satellite time and frequency
transfer (TWSTFT) was successfully demonstrated 
over a very long baseline of 9,000 km, 
established between the National Institute of Information and
Communications Technology (NICT) and the Physikalisch-Technische 
Bundesanstalt (PTB). 
We verified that the carrier-phase TWSTFT (TWCP) result 
agreed with those obtained by 
conventional TWSTFT and GPS carrier-phase (GPSCP) techniques. 
Moreover, a much improved short-term instability for 
frequency transfer of $2\times10^{-13}$ 
at 1 s was achieved, which is at the same level as previously confirmed 
over a shorter baseline within Japan. 
The precision achieved was so high that 
the effects of ionospheric delay became significant 
which are ignored in conventional TWSTFT even over a long link. 
We compensated for these effects using ionospheric delays computed from 
regional vertical total electron content maps. The agreement between the 
TWCP and GPSCP results 
was improved because of this compensation. 

\maketitle

\section{Introduction}

There is a need for synchronization between remote clocks in radio
astronomy, particle accelerators, and metrology \cite{BauchPeik, Peik, Frisch}. 
With the development 
of highly accurate optical clocks and an increase in the number of
simultaneously controlled instruments, there has been 
a corresponding increase in the required accuracy for time 
and frequency transfer. Recently, there 
have been extensive studies regarding time and frequency transfer 
over optical fibers \cite{Poland, Rost, Lopez}. 
Time synchronization based on an 
Ethernet-based network has been developed to realize sub nanosecond 
accuracy \cite{WR}. 
Moreover, optical frequency dissemination 
through optical fiber links enables the comparison of remote optical clocks and 
the measurement of the absolute transition frequency with reference to 
a remote cesium fountain clock without any limitations with respect 
to its accuracy \cite{Yamaguchi, Matveev}. 
The transfer length exceeded 900 km \cite{Science_920km}. \\
\hspace{1em}
Optical fiber links may be established for 
time and frequency transfer among various institutes 
located within the same continental region. However, 
it will be challenging to establish intercontinental
optical fiber links. To make overseas connections, the utilization of 
satellite links such as GPS and two-way satellite time and frequency
transfer (TWSTFT) will be indispensable, 
and there is a need to improve their measurement precision and accuracy 
\cite{Fujieda1, PiesterSchnatz}. 
These techniques have been adopted since 1999 for the determination of 
Coordinate Universal Time (UTC) and 
International Atomic Time (TAI) \cite{BIPM}. 
Practically all institutes that maintain time and frequency standards 
adopt GPS time and frequency transfer 
to become part of the network of institutes collaborating in the 
realization of TAI with the BIPM. 
The use of GLONASS and Galileo has become more and more common practice, 
but in this paper 
%we restrict on GPS frequency transfer as this was used
we limit ourselves to use GPS frequency transfer 
in parallel to carrier-phase TWSTFT (TWCP). 
In addition, some institutes also use TWSTFT as 
a method that is independent of GPS. 
Unfortunately, the measurement precision has not significantly 
improved since its implementation many years ago \cite{Kirchner}. 
This is mainly due to the high costs for the lease of large
bandwidths of transponder capacity on communication satellites. \\
\hspace{1em}
Both TWSTFT and GPS utilize phase-modulated signals with pseudo-random
noise codes, and 
the code phase has been used to measure propagation time. 
The Coarse/Acquisition (C/A) code rate and carrier frequency of 
the GPS L1 signal are 1.023 Mega chip-per-second (cps) and 1.57542 GHz, 
respectively \cite{GPS_geodesy}. 
The measurement precision realized by the C/A code is about 10 ns with an
averaging time of 13 min for time transfer. 
With the implementation of GPS carrier-phase measurements (GPSCP), the 
precision was improved to several tens of picoseconds. 
On the other hand, a code rate of 2.5 Mcps 
is typically used in
TWSTFT. The measurement precision (jitter) for 
1-second data points 
%secondly recorded data 
is about 0.5 ns when taken with a carrier-to-noise density ratio of 55 dBHz 
\cite{Piester2008}. 
In the case where the carrier-phase measurement 
is implemented at about 10 GHz, we can expect 
the measurement precision to be 
improved by three orders of magnitude. 
%Fonville et al. introduced the 
%use of the carrier phase to TWSTFT \cite{Fonville}. 
Sch\"{a}fer et al. introduced the use of the carrier phase to TWSTFT
\cite{Schaefer} and Fonville et al. demonstrated an intercontinental test
result for a few hours \cite{Fonville}. 
The National Institute of Information and Communications Technology
(NICT) confirmed that the result obtained using 
TWCP agreed with that of GPSCP 
over a short baseline of 100 km \cite{Fujieda1}. 
After this regional exercise, we performed a more ambitious 
TWCP experiment over one of the longest
baselines worldwide with the intention to evaluate 
the error sources that would be otherwise negligible 
over a short baseline. Our target is to improve the measurement
precision of TWSTFT in intercontinental links. 
This study covers availability, error sources, and 
uncertainty over a longer baseline for TWCP. First, we introduce 
the time difference using carrier-phase information. Next, we discuss 
the experimental equipment, results, and discussions. Finally, we 
present our conclusions. 

\section{Description of carrier-phase two-way satellite frequency transfer}

Two Earth stations, A and B, work as a pair. 
Four sets of phase information are required to describe the time difference by
TWCP using a commercial geostationary satellite 
because we have four unknown values: the time difference between the two
stations, the phase fluctuation induced by the translation of the received
to the transmitted signal at the satellite, and two 
geometrical distances 
between the satellite and the two stations involved. 
Each station transmits a signal and detects the phases of 
its own transmitted signal and that of the signal received 
from the counterpart station. \\
\hspace{1em}
Uplink and downlink signal angular frequencies are expressed as $\omega_{u}$
and $\omega_{d}$, respectively; times of the reference clocks at station
A and B are expressed as $\tau_{a}(t)$ and $\tau_{b}(t)$ in seconds, 
respectively; 
angular frequency of the translation onboard oscillator is expressed as
$\omega_{s}=\omega_{u}-\omega_{d}$; 
time of the onboard clock is expressed as $\tau_{s}(t)$ in seconds. 
When a transmitted signal from station A,
$sin(\omega_{u}t+\omega_{u}\tau_{a}(t))$, arrives at the satellite, 
it shifts to 
$sin(\omega_{u}^{'}t+\omega_{u}\tau_{a}(t))$ because of the Doppler
effect. 
Here $\omega_{u}^{'}$ is given with the radial velocity $v_{a}(t)$ from station A to
the satellite as
\begin{eqnarray}
 \omega_{u}^{'}&=&\omega_{u}(1-\frac{v_{a}(t)}{c}) 
\end{eqnarray}
Here c is the speed of light. The signal is down-converted by the onboard signal
$sin(\omega_{s}t+\omega_{s}\tau_{s}(t))$ to
$sin[(\omega_{u^{'}}-\omega_{s})t+\omega_{u}\tau_{a}(t)-\omega_{s}\tau_{s}(t)]$. When
this signal arrives at station B, it shifts to 
$sin(\omega_{d}^{''}t+\omega_{u}\tau_{a}(t)-\omega_{s}\tau_{s}(t))$ because
of the Doppler effect. 
Using a local signal $sin(\omega_{d}t+\omega_{d}\tau_{b}(t))$ at station B, the received signal is finally down-converted to
$sin[(\omega_{d}^{''}-\omega_{d})t+\omega_{u}\tau_{a}(t)-\omega_{s}\tau_{s}(t)-\omega_{d}\tau_{b}(t)]$. 
The angular frequency received at station B with the Doppler shift is
expressed with the radial velocity $v_{b}(t)$ from station B to the
satellite as 
\begin{eqnarray}
 \omega_{d}^{''}&=&(\omega_{u}^{'}-\omega_{s})(1-\frac{v_{b}(t)}{c})=\omega_{d}-\omega_{d}\frac{v_{b}(t)}{c}-\omega_{u}\frac{v_{a}(t)}{c}+\omega_{u}\frac{v_{a}(t)}{c}\frac{v_{b}(t)}{c}
\end{eqnarray}
In the case of a geostationary satellite, the quantity of $v(t)/c$ is 
in the range of 
$10^{-8}$ to $10^{-9}$. Thus, it is thought that the fourth term on the
right hand is less
than $10^{-16}$ and negligible. To simplify the description, 
it is eliminated hereafter. 
The phase information of the signal 
%which is transmitted from station A and received at station B 
includes the ionospheric and tropospheric delays 
along the signal path multiplied by 
the signal angular frequency including
the Doppler shift. Considering the small magnitude of $v(t)/c$ 
we neglect the effects of the Doppler shifts in the ionospheric and
tropospheric delays. 
Accordingly, the phase information $\phi_{ab}(t)$ in radian of the signal
which is transmitted from station A and received at station B is given by 
\begin{eqnarray}
\phi_{ab}(t) &=&
 \omega_{u}\tau_{a}(t)-\omega_{s}\tau_{s}(t)-\omega_{d}\tau_{b}(t)-(\omega_{u}v_{a}(t)t+\omega_{d}v_{b}(t)t)/c
 \nonumber \\ 
&
 &+\omega_{u}^{'}I_{ua}(t)+\omega_{d}^{''}I_{db}(t)+\omega_{u}^{'}T_{a}(t)+\omega_{d}^{''}T_{b}(t)+\omega_{u}\frac{v_{a}(t)}{c}\frac{v_{b}(t)}{c}t
\nonumber \\
&\approx&
 \omega_{u}\tau_{a}(t)-\omega_{s}\tau_{s}(t)-\omega_{d}\tau_{b}(t)-(\omega_{u}\rho_{as}(t)+\omega_{d}\rho_{bs}(t))/c \nonumber\\
& & +\omega_{u}I_{ua}(t)+\omega_{d}I_{db}(t)+\omega_{u}T_{a}(t)+\omega_{d}T_{b}(t)
\end{eqnarray}
Here $\rho_{as}$ and $\rho_{bs}$ are the geometrical distances 
between the satellite and stations A and B, respectively. 
$T_{a}$ and $T_{b}$ are tropospheric delays in seconds which are
frequency independent 
over stations A and B, respectively. 
$I_{ij}$ is the ionospheric delay correction term in seconds with frequency $f_{i}$ 
$(i = u$ or $d)$ 
at position j which can be represented as follows: 
\begin{equation}
I_{ij}(t) = \frac{40.3\cdot TEC_{j}(t)}{c\cdot f_{i}^{2}} 
\end{equation}
$TEC_{j}(t)$ is the total electron content, the total
number of free electrons along the signal path 
at position j, which is conventionally measured in TEC units, 
1 TECU = $10^{16}$ electrons/$m^{2}$.
$f_{u}$ and $f_{d}$ are the uplink and downlink frequencies,
respectively. 

Similarly, the phase information  
from B to A, from A to A, and from B to B, are respectively represented by
\begin{eqnarray}
\phi_{ba}(t) &=&
 \omega_{u}\tau_{b}(t)-\omega_{s}\tau_{s}(t)-\omega_{d}\tau_{a}(t)-(\omega_{u}\rho_{bs}(t)+\omega_{d}\rho_{as}(t))/c \nonumber\\
& & +\omega_{u}I_{ub}(t)+\omega_{d}I_{da}(t)+\omega_{u}T_{b}(t)+\omega_{d}T_{a}(t)\\
\phi_{aa}(t) &=&
 \omega_{u}\tau_{a}(t)-\omega_{s}\tau_{s}(t)-\omega_{d}\tau_{a}(t)-(\omega_{u}\rho_{as}(t)+\omega_{d}\rho_{as}(t))/c\nonumber\\
& & +\omega_{u}I_{ua}(t)+\omega_{d}I_{da}(t)+\omega_{u}T_{a}(t)+\omega_{d}T_{a}(t)\\
\phi_{bb}(t) &=&
 \omega_{u}\tau_{b}(t)-\omega_{s}\tau_{s}(t)-\omega_{d}\tau_{b}(t)-(\omega_{u}\rho_{bs}(t)+\omega_{d}\rho_{bs}(t))/c\nonumber\\
& & +\omega_{u}I_{ub}(t)+\omega_{d}I_{db}(t)+\omega_{u}T_{b}(t)+\omega_{d}T_{b}(t)
\end{eqnarray}

In general, the geometrical distance from station A to the satellite and the
corresponding distance from station B to the satellite 
are not symmetrical. As a result, 
there is a difference in 
the arrival times of signals from stations A and B at the satellite. 
This difference is generally a few milliseconds \cite{Tom}. 
We assume that the onboard translation oscillator signal is stable 
during this time difference, and the induced phase jitter is negligible 
and thus consider only the one unknown \(\tau_{s}\) in all equations
(3), (5) - (7). 

By subtracting equation (3) from equation (5), 
\begin{eqnarray}
\phi_{ab}(t)&-&\phi_{ba}(t) = 
\omega_{+}(\tau_{a}(t)-\tau_{b}(t))-\omega_{-}(\rho_{as}(t)-\rho_{bs}(t))/c\nonumber\\
&+&\omega_{-}(T_{a}(t)-T_{b}(t))+\omega_{u}(I_{ua}(t)-I_{ub}(t))
-\omega_{d}(I_{da}(t)-I_{db}(t)) ,
\end{eqnarray}
and (6) and (7), 
\begin{eqnarray}
\phi_{aa}(t)&-&\phi_{bb}(t) = 
\omega_{-}(\tau_{a}(t)-\tau_{b}(t))-\omega_{+}(\rho_{as}(t)-\rho_{bs}(t))/c\nonumber\\
&+&\omega_{+}(T_{a}(t)-T_{b}(t))+\omega_{u}(I_{ua}(t)-I_{ub}(t))
+\omega_{d}(I_{da}(t)-I_{db}(t)) ,
\end{eqnarray}
the terms related to the onboard signal are canceled. 
Here, $\omega_{+}\equiv \omega_{u}+\omega_{d}$ and 
$\omega_{-}\equiv \omega_{u}-\omega_{d}$ apply. 

In TWCP, the geometrical distances in equations (8) and (9), 
$\rho_{as}$ and $\rho_{bs}$, are unknown. Therefore, 
it is necessary to further subtract these intermediate results 
to cancel them. 
\begin{eqnarray}
\omega_{+}(\phi_{ab}(t)-\phi_{ba}(t))&-&\omega_{-}(\phi_{aa}(t)-\phi_{bb}(t))
\nonumber\\
&=&(\omega_{+}^{2}-\omega_{-}^{2})(\tau_{a}(t)-\tau_{b}(t))\nonumber\\
&+&2\omega_{u}\omega_{d}[(I_{ua}(t)-I_{ub}(t))-(I_{da}(t)-I_{db}(t))]
\end{eqnarray}
The time difference between stations A and B is finally computed as follows: 
\begin{eqnarray}
\tau_{a}(t) - \tau_{b}(t) &=&
 \frac{\omega_{+}\alpha(t)-\omega_{-}\beta(t)}{\omega_{+}^{2}-\omega_{-}^{2}}\nonumber\\
 & &+\frac{2\omega_{u}\omega_{d}}{\omega_{+}^{2}-\omega_{-}^{2}}[(I_{da}(t)-I_{ua}(t))-(I_{db}(t)-I_{ub}(t))]\\
\alpha(t) &\equiv& \phi_{ab}(t) - \phi_{ba}(t) \nonumber \\
\beta(t) &\equiv& \phi_{aa}(t) - \phi_{bb}(t) \nonumber %\\
\end{eqnarray}

Here the tropospheric delays are canceled. On the right hand of equation (11), 
the first term is derived from the four observed phases, and 
the second one is for ionospheric delay correction. 
Typically, TWSTFT uses uplink and downlink frequencies of 
14 GHz and 11 GHz, respectively. In this case, the coefficients of
$\omega_{+}/(\omega_{+}^{2}-\omega_{-}^{2})$ and 
$\omega_{-}/(\omega_{+}^{2}-\omega_{-}^{2})$ become 
$7\times10^{-12}$ and $8\times10^{-13}$, respectively. 
Thus, a measurement precision of $10^{-13}$ seconds can be expected
when the phase detection has a resolution of less than \\0.1 radians. 
\\ \hspace{1em}
When $TEC_{j}(t)$ is obtained by other methods such as a theoretical model
or actual measurement, the ionospheric correction term can be
calculated. The ionospheric delay needs to be considered carefully because of 
the frequency difference between 
uplink and downlink and the $TEC$ difference between the ionospheres 
at stations A and B. 
The TWCP frequencies are, however, higher than those used in GPS such
that the effect due to the ionosphere (prop. 1/\(f^{2}\)) is 
smaller than that encountered in GPS time transfer. 
The amplitude could be up to a few hundred picoseconds, and is assumed to be
negligible in conventional TWSTFT (hereafter TWCode), 
but it cannot be neglected in TWCP. 

\section{Experimental setup}

\begin{figure}[htb]
\begin{center}
\resizebox{10cm}{!}{\includegraphics{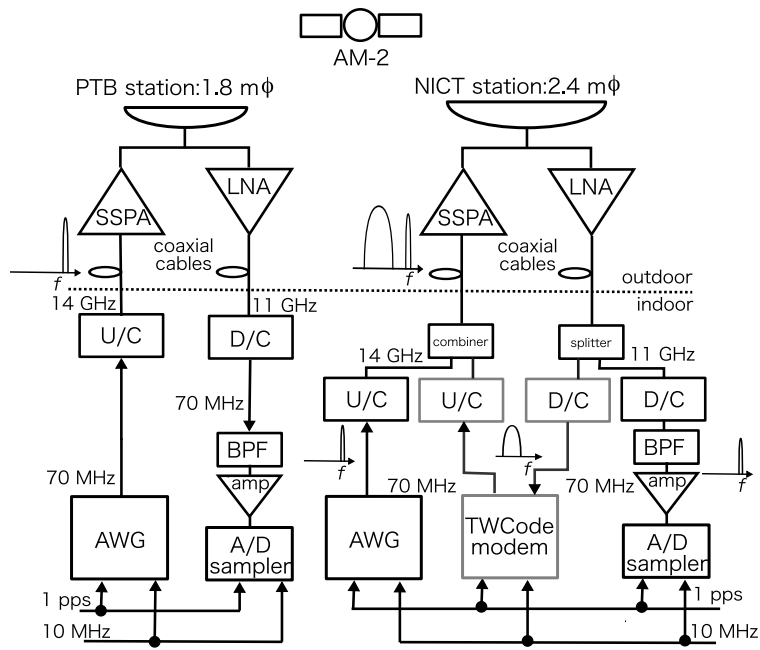}}
\caption{Earth station configuration for NICT-PTB TWCP measurement. 
U/C: frequency up-converter, D/C: frequency
 down-converter, SSPA: solid-state power amplifier, LNA: low-noise
 amplifier, BPF: band-pass filter, amp: amplifier, AWG: arbitrary
 waveform generator, A/D sampler: analog-to-digital sampler. 
The instruments in gray color were not used in the TWCP measurement. }
\label{fig:ES_setup}
\end{center}
\end{figure}

The TWCP experiment was performed using a link 
between NICT and Physikalisch-Technische
Bundesanstalt (PTB) by employing the geostationary 
communication satellite AM-2 located at a longitude of 80 degrees East. 
The actual distance between the two sites is approximately 9,000 km, 
making it one of the longest baselines in the world. 
In parallel, TWCode measurement was performed once every hour 
using the same satellite that has been used to link 
Asian and European laboratories belonging to the TAI networks
\cite{BIPM}. 
The satellite IS-4 was used for this purpose for many years. 
Due to its malfunction four years ago, however, it was replaced by the
satellite AM-2 to continue the link. 
The TWCP measurement was performed using a narrow
frequency band with a bandwidth of 200 kHz, whose center frequency is 
2 MHz higher than the frequency band of 2.5 MHz used for 
TWCode measurements. 
Because of limited transponder working time, the measurements were 
continuously performed only from 12:00 UTC to 22:00 UTC every day. 
The frequency parameters are listed in Table \ref{tbl:AM2}. 
\begin{table}
\caption{Link parameters.}
\label{tbl:AM2}
\begin{indented}
\item[]\begin{tabular}{@{}lll}
\br
Parameters & TWCP & TWCode\\
\mr
Uplink center frequency [MHz] & 14262 & 14260 \\
Downlink center frequency [MHz] & 10962 & 10960 \\
Bandwidth [MHz] & 0.2 & 2.5 \\
\br
\end{tabular}
\end{indented}
\end{table}

The experimental setup is shown in Figure \ref{fig:ES_setup}. 
In each station we employed a narrow-band pseudo-random noise
signal to reduce the rental fee for the satellite transponder and 
to avoid effects due to fading or interference between the signals 
transmitted simultaneously from the two sites. 
To generate such a signal, we used 
an arbitrary waveform generator (AWG) developed for 
a dual pseudo-random-noise experiment \cite{Amagai, Gotoh}. 
Pseudo-random noise of 127.75-kcps, which consisted of a 
maximum-length-sequence code 
with a length of 511 bits, was generated by the AWG whose output  
bandwidth was limited by a 200-kHz digital filter. 
%The prepared code pattern was uploaded to the AWG via a USB port. 
%The AWG periodically converted the given samples 
%to an analog signal with a sampling rate of 204.6 MHz 
%and a dynamic range of 8 bits 
%The AWG worked synchronizing the external reference 
%signals of 10 MHz and 1 pps. 
In addition, it had a feature to overlay 50 bps data, allowing it 
to identify the start of a code pattern which is 
synchronized with the external 1 pps, and 
to reduce the interference between codes.  
A signal with a center frequency of 70 MHz, which was generated by the
AWG, was then up-converted to a higher 14 GHz frequency using a 
frequency up-converter (U/C). The signal was amplified by a
solid-state power amplifier (SSPA) and fed to the antenna. 
Both stations used a different code in their transmission path. 

The signal received by the antenna with a center frequency of 11 GHz 
was amplified by a low-noise amplifier (LNA) and 
down-converted to 70 MHz by a frequency down-converter (D/C). 
The U/C and D/C were phase locked to external 10-MHz signals. 
The signal passed through a band-pass filter (BPF) with a frequency 
bandwidth of \\2 MHz, 
and it was amplified by an additional amplifier. Then, it was sent 
to an analog-to-digital (A/D) sampler. 
The A/D sampler was originally 
developed for very long-baseline interferometry (VLBI) \cite{VLBI}. 
Data sampling was performed with a sampling rate of 64 MHz 
in synchronized timing with the external
reference signals of 10 MHz and 1 pps. 
%The sampled data was decimated to 1/8 after filtering by a digital
%filter to reduce the load to the computer, 
%and the data were then transmitted to the computer via USB. 
After splitting the sampled data into in-phase and quadrature components, 
the cross correlation with a replica code, group delay, 
and carrier-phase detection were sequentially conducted in the
computer. 
Each antenna received two pseudo-random noise signals, that is, the own and the
one transmitted from the counterpart station. Therefore two 
data-processing sequences ran in parallel. 
The code and carrier-phase information was then written every 20 ms.
Finally, the midpoint of the second-order fit 
to the 50 data points obtained during each second was 
defined as the data at that second. 
With the exception of the AWG and A/D sampler, the instruments 
were identical to those used for TWCode stations.\\
\hspace{1em}
The PTB station was dedicated for conducting TWCP measurements; it was 
equipped with a 1.8-m antenna dish. 
%, where an SSPA, LNA, and the antenna 
%were installed on the roof-top. 
The elevation angle was 3.7$^{\circ}$. 
%The other instruments were installed indoors. Two coaxial cables 
%connected the indoors to outdoors, where the signals with Ku-band 
%frequencies were passed through. 
The PTB station was connected to UTC(PTB) signals of 1 pps and 10 MHz. 
UTC(PTB) is derived from a steered active hydrogen maser \cite{Bauch2} 
and the frequency stability is at the \(10^{-15}\) level for averaging times 
exceeding \(10^{4}\) seconds. 
Because the PTB station was located a
few hundred meters away from the building in which hydrogen masers are
operated 
and the UTC(PTB) time scale is realized, the reference signals of 10 MHz and
1 pps were distributed using a 1-km-long optical fiber link. 
It was confirmed that the distribution instability was sufficiently 
small compared to the stability of the reference signal 
without any fiber-length stabilization. 
\\ 
\hspace{1em}
On the other hand, the NICT station with a 2.4-m dish 
was shared by the TWCode and TWCP measurements.  
The elevation angle was 16.0$^{\circ}$, 
and both measurements were performed simultaneously. 
%The antenna, SSPA, and LNA were installed on the roof-top. 
%Only the LNA at the NICT station was installed in the 
%temperature-controlled box. 
%The U/C, D/C, AWG, A/D sampler, and others were done on the inside. 
The 200-kHz and 2.5-MHz wide signals at 
sufficiently different carrier frequencies were generated 
by the AWG and a TWCode modem, respectively, and were combined 
after the frequency up-conversion from 70 MHz to 14 GHz to avoid the 
generation of inter-modulation. 
%, and they were then transmitted outdoors
% by a coaxial cable, after which they were 
%amplified by the SSPA. 
The received signal was split into two components
after amplification by the LNA, and the two components were input 
into different D/Cs. 
The 200-kHz signal was separated from the 2.5-MHz
TWCode signal by the BPF. We confirmed that there was no degradation in the
measurement precision caused by the combined processing of the two signals. 
The NICT station was connected to UTC(NICT) signals of 1 pps and 10
MHz. UTC(NICT) is generated from a steered active hydrogen maser and the
frequency stability is at the \(10^{-15}\) level for averaging times
exceeding \(10^{4}\) seconds \cite{Nakagawa}. 

\section{Error sources over a very long baseline}

\subsection{Compensation for Ionospheric delay}

\begin{figure}[htb]
\begin{center}
\resizebox{13cm}{!}{\includegraphics{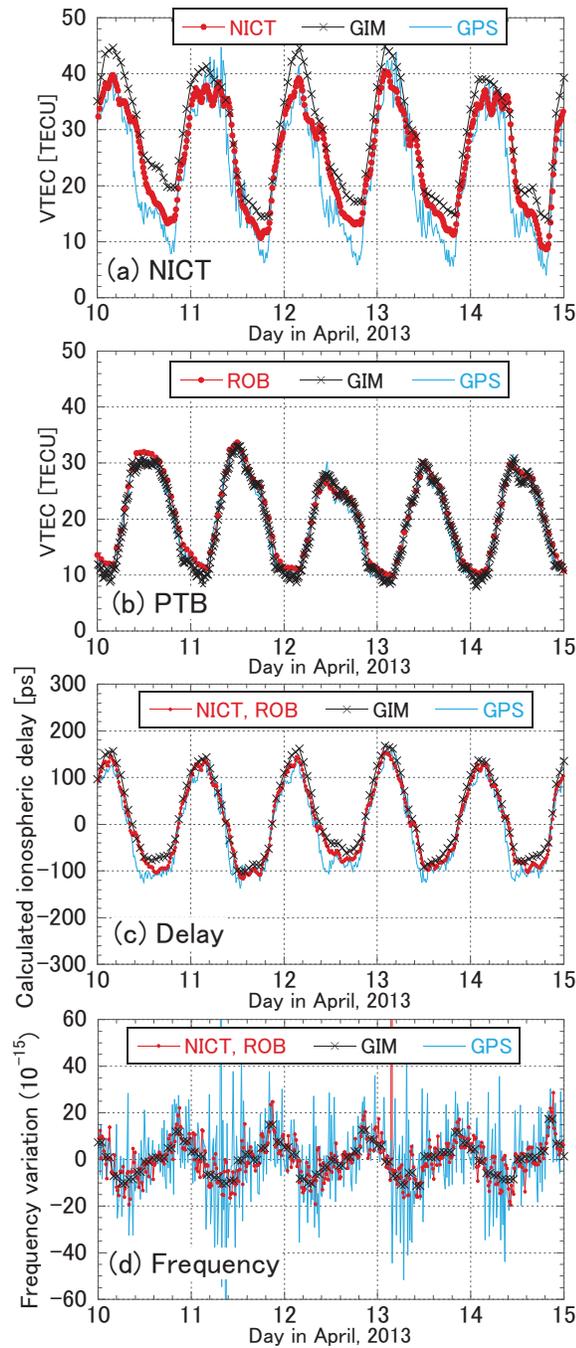}}
\vspace{-6mm}
\caption{(a) VTEC over NICT computed from the NICT VTEC map, GIM, and
 GPS, 
(b) VTEC over PTB computed from the ROB VTEC map, GIM, and GPS, 
(c) calculated ionospheric delay and (d) expected 
frequency variation in the NICT-PTB link computed from the 
NICT and ROB VTEC maps, GIM and GPS.}
\label{fig:tec_map}
\end{center}
\end{figure}

Because ionospheric delay is inversely proportional to the square of the
signal frequency, it is not canceled out in TWSTFT because of the frequency
difference between the uplink and downlink. In TWCode, 
it is assumed that the impact of the 
ionospheric delay is negligible \cite{Bauch1}. 
On the other hand, it should be considerable in the TWCP link 
taking the high performance into account. 
Because the TWCP measurement is carried out by using a 
pair of uplink and downlink frequencies, the TEC 
along the signal path cannot be measured; 
two downlink frequencies would be needed as it is 
the case when a dual-frequency GPS receiver is employed. 
%The ionospheric delay is derived from the actual measurement. 
We read out the vertical total electron content
(VTEC) over NICT and PTB from some VTEC maps and 
compensated for the ionospheric delays present in the TWCP result.  
The Center for Orbit 
Determination in Europe (CODE) \cite{CODE} generates the Global
Ionosphere Map (GIM) based on observations that are made 
at about 150 GPS sites around the world. 
The GIM provides VTEC maps with a time resolution 
of 2 h and latitude/longitude resolution of 2.5/5.0 degrees. 
It is assumed that the VTEC is present in an infinitely thin layer 
at a height of 450 km. 
To compute the VTEC $E$ at latitude $\beta$, longitude $\lambda$, and
universal time $t$, we first read out four VTEC values, 
$E_{0,0}$, $E_{1,0}$, $E_{0,1}$, $E_{1,1}$, 
at epoch $T_{i}$ around the coordinates
$(\lambda, \beta)$ and perform grid interpolation:\\
\begin{eqnarray}
E_{i}(\beta,\lambda)&=&
E_{i}(\lambda_{0}+p\Delta\lambda,
\beta_{0}+q\Delta\beta)\nonumber \\
&=&(1-p)(1-q)E_{0,0}+p(1-q)E_{1,0}+q(1-p)E_{0,1}+pqE_{1,1},
\end{eqnarray}
where $0\leq p < 1$ and $0\leq q < 1$. $\Delta\lambda$ and
$\Delta\beta$ are the grid resolutions in longitude and latitude, respectively. 
Then, we perform the interpolation of time using 
the consecutive VTEC maps:\\
\begin{eqnarray}
E(\beta,\lambda,t)=\frac{T_{i+1}-t}{T_{i+1}-T_{i}}E_{i}(\beta,\lambda)+\frac{t-T_{i}}{T_{i+1}-T_{i}}E_{i+1}(\beta,\lambda),
 %\nonumber
\end{eqnarray}
where $T_{i}\leq t < T_{i+1}$. 
Next, we compute a slant TEC along the signal path from VTEC
$E(\beta,\lambda,t)$ using the elevation angle $z$ of an Earth
station to a communication satellite. 
\begin{eqnarray}
z^{'} &=& sin^{-1}[\frac{Rsin(\pi/2-z)}{R+h}],\\
TEC(\beta,\lambda, t) &=& \frac{E(\beta,\lambda,t)}{cos z^{'}},%\nonumber
\end{eqnarray}
where $R$ is the radius of the Earth and $h$ is the height of the
ionosphere \cite{GPS_geodesy}. 
Thus, the ionospheric delay over an Earth station in TWSTFT 
is derived using equation (4) and $TEC(\beta,\lambda,t)$ 
computed from a VTEC map \cite{ionex}. 
However, the time resolution and grid interval of the GIM 
are not sufficiently dense considering the TWCP measurement 
rate of every second. Therefore, we 
adopted regional VTEC maps provided by the Royal Observatory of Belgium 
(ROB) \cite{ROB} and NICT \cite{NICT_TEC} over Europe and Japan,
respectively. 
The ROB VTEC map is based on real-time GPS
observations obtained from more than 100 sites belonging to the 
EUREF Permanent Network \cite{EUREF}. 
The grid interval is ($0.5^{\circ},
0.5^{\circ}$), and the time resolution is 15 min. 
On the other hand, the NICT VTEC map is 
generated using the GPS Earth Observation Network (GEONET) data from
about 200 sites in Japan \cite{GEONET}, and its grid interval is ($2.0^{\circ},
2.0^{\circ}$), while its time resolution is also 15 min. \\
\hspace{1em}
We first evaluated the disagreement between NICT and ROB VTEC maps, 
and the GIM. In addition, the VTEC values were calculated from the
CGGTTS data \cite{CGGTTS} of dual-frequency GPS receivers at NICT and PTB 
using equation (4), (14) and (15). 
Since the GPS receivers have multi channels, the mean values of each
measurement epoch every 16 min were calculated. 
Figures \ref{fig:tec_map} (a) and (b) show the VTEC values over NICT and
PTB, which were derived from NICT and ROB VTEC maps, GIM and GPS data. 
Time and grid interpolation was performed 
as previously described. The time resolutions 
were every 15 min, 1 h and 16 min for the results using NICT and ROB
VTEC maps, GIM, and GPS, respectively. 
We found that the disagreements between the NICT VTEC
map and GIM was always less than 6 TECU. 
In addition, 
the maximum difference between the NICT VTEC map and the result by 
GPS was 7 TECU. 
In particular, the ROB VTEC map shows a 
good agreement with GIM and the result by GPS. 
Thus the regional VTEC maps were considered appropriate for our purpose 
and were employed for compensation because they had a minimum time resolution. 
Figure \ref{fig:tec_map} (c) depicts the calculated ionospheric delays   
described in equation (11) and Figure \ref{fig:tec_map} (d) shows 
the converted frequency variations whose data spacing is 15 min. 
The result by GPS shows a larger frequency deviation, however,  
the mean value looks consistent with those by GIM and the regional VTEC maps.  
Depending on the VTEC maps utilized, no significant 
differences were observed. 
The amplitudes exceeding 100 ps are caused by amplification of 
the effect due to 
the local time difference between Japan and Germany 
and the low elevation angle 
at the PTB station. However, 
such a local time difference and low elevation angle are unavoidable 
over a very long baseline. 

\subsection{Other sources}

\begin{table}[htb]
%\caption{Error sources over a very long TWCP link.}
\caption{Delay sources over a very long TWCP link.}
\label{tbl:error}
\begin{indented}
\item[]\begin{tabular}{@{}ll}
\br
Item & Amplitude in time [ps]\\
\mr
phase jitter in satellite transponder & unknown \\
phase jitter in instruments & 0.2 (frequency converter)\\
phase variation in instruments & 100 \\%700 ps/K (frequency converter)\\
ionosphere & 100 \\
troposphere & 3 \\
Sagnac effect & 10\ to\  150 \\
2nd order Doppler shift & 2 \\
\br
\end{tabular}
\end{indented}
\end{table}

The possible %error 
delay sources over a very long baseline are 
listed in Table \ref{tbl:error}. 
With the exception of ionospheric delay, these error sources 
are roughly estimated in this subsection. 
In our description of the time difference using carrier-phase information, 
it is assumed that the induced phase jitter during frequency
conversion at the satellite transponder is negligible; however, 
the impact is actually unknown. 
%It seems that it is less than the phase
%jitter of 0.1 ps that was achieved in a common-clock measurement as
%shown in Figure \ref{fig:mdev_cc}. 
We previously evaluated phase jitters caused by instruments
\cite{Fujieda1}. 
From our results, it was determined that a jitter of 0.2 ps caused by 
a commercially available frequency converter is the largest in the TWCP
system. The temperature coefficients of the instruments were also
measured using a temperature-controlled bath. We confirmed that 
a commercially available frequency converter had the largest value of
about $\pm$700 ps/K. As shown in the next section, outdoor instruments
can also cause phase variations. 
The phase variation caused by a temperature
variation would be considerable, depending on magnitude and period. 
We estimate that this effect to be 200 ps peak-to-peak at maximum.\\
\hspace{1em}
The tropospheric delay is normally assumed to be frequency
independent up to frequencies slightly above 10 GHz, 
and it is believed that it is canceled out in TWSTFT. However, in
principle, the troposphere refraction index is dependent on frequency, and 
it induces a small impact in TWSTFT. As shown in equation (11), 
the delays that are dependent on the frequency are coupled with the ionosphere
delays, and they induce phase variation. 
Piester et al. estimated a peak-to-peak amplitude of 2 ps 
for the NICT-PTB link in the case involving IS-4 \cite{Piester1}. 
Considering the difference in the elevation 
angles, it may attain a peak-to-peak of 5 ps 
in the case involving the AM-2 satellite currently used. 
When there is such a delay
variation within a 24-h period, it is 
equivalent to an increase of the Allan deviation by $10^{-16}$ at half day 
averaging time calculated as $5\times10^{-12}$ s/$43200$ s. 
This magnitude can still be considered to be smaller than other error
sources. \\
\hspace{1em}
The Sagnac effect is caused by the rotation of the Earth and it 
induces a propagation time difference between two stations. 
TWSTFT is generally performed using a geostationary satellite. 
A delay correction is routinely done in TWCode, 
assuming that the propagation time difference is constant. 
Because of slight movements around the central position of the satellite, 
the difference in the propagation time, however, 
has a periodic daily variation. 
The amplitude of the time difference depends on the position of
the Earth station and the magnitude of the movement of the satellite. 
We computed the satellite position using the two line elements (TLE) 
published by the North American Aerospace Defense Command (NORAD)
\cite{NORAD} and estimated the time difference due to the Sagnac
effect. 
The amplitudes of the time difference due to the Sagnac effect 
in the NICT-PTB link obtained using the IS-4 satellite 
varied from several hundred picoseconds to several tens of picoseconds, 
depending on the satellite's orbit. 
On the other hand, amplitudes of only some tens of picoseconds were
computed in the case involving the AM-2 satellite. 
Evidently, the position of the AM-2 satellite is more tightly 
controlled than that of the IS-4 used to be. 
Here, we do not take this correction into account. \\
\hspace{1em}
In equation (3), the term of the second order Doppler shift 
$v_{a}/c \cdot v_{b}/c$ is negligible. 
We measured the magnitudes of $v/c$ using the IS-4
and AM-2 satellites, which had a maximum value of $10^{-9}$ and 
less than $10^{-8}$, respectively. 
Thus, it is thought that the term induces a maximum instability
of $10^{-16}$ with a diurnal pattern. Accordingly, we 
estimated an amplitude of 2 ps, and we do not consider a correction. 

\subsection{Environmental effect}

\begin{figure}[htb]
\begin{center}
\resizebox{12cm}{!}{\includegraphics{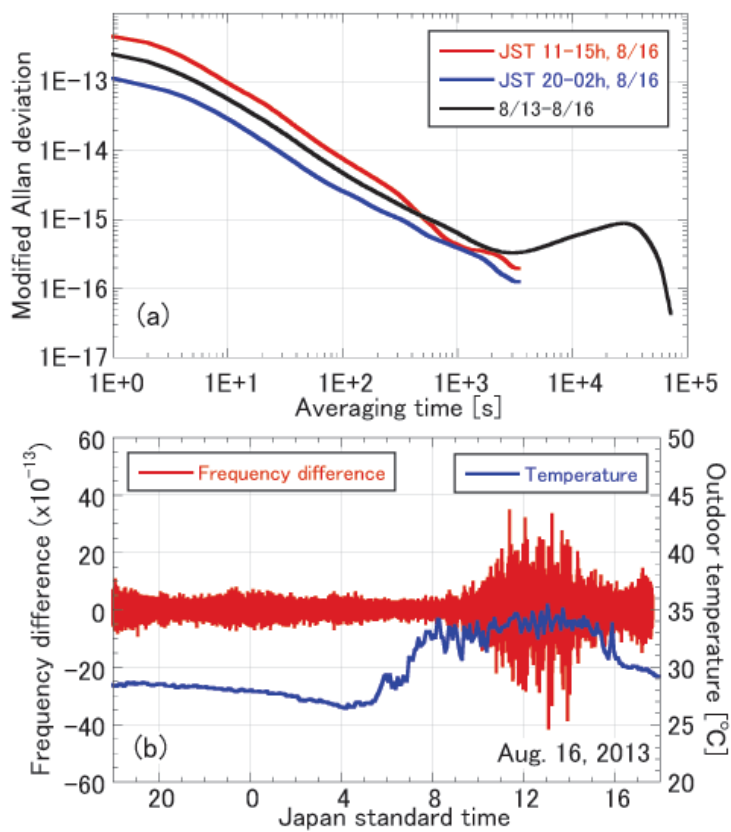}}
\vspace{-58mm}
\caption{(a) TWCP instabilities and (b) frequency difference with
 the outdoor temperature in the common clock measurement at NICT. }
\label{fig:mdev_cc}
\end{center}
\end{figure}

During TWCP measurements, we observed some phenomena 
that may have been related to the outdoor instruments and temperature. 
Transmission power decreased by about 10 dB 
when the outdoor temperature at PTB increased to more than 30 $^{\circ}$C. 
The measurement precision of TWCP degraded by a factor of 4 during the
daytime, as shown in Figure \ref{fig:mdev_cc}, 
when the outdoor temperature was above 30 $^{\circ}$C at NICT. 
There were little changes in the transmission power. 
The former occurred in June 2013 at PTB, while the latter occurred 
during the common clock measurement using two antenna dishes 
performed at NICT in August 2013. 
We checked the output power levels of the indoor instruments and
confirmed that they had not changed in both cases.\\ 
\hspace{1em}
For the former case, one possible explanation is that the SSPA (see Figure 1) 
was affected by the heat, 
and the gain subsequently decreased. 
After the SSPA was shielded 
from direct sunlight, the gain
recovered within a few hours. Phase variations were also observed. \\
\hspace{1em}
For the common clock measurement in the latter case, 
in particular, there was a daily variation 
with an amplitude of about 20 ps, and it had a correlation with
the variation of the outdoor temperature. 
Besides, the short-term instability observed during the daytime was clearly
degraded compared to that in the nighttime, as shown in Figure
\ref{fig:mdev_cc} (a). 
Because the LNA was temperature-stabilized and the indoor temperature did
not vary much at the NICT station, it is thought that 
the outdoor temperature rise caused the phase jitter to increase. 
We concluded that more attention should be paid to the
condition of the outdoor instruments. In particular, in the case of a 
very long baseline, the phase variation attributed to the temperature
variation may be enlarged by the local time difference. 
It would be required to 
avoid heating and temperature variations on the installed equipment 
to fully utilize the performance of TWCP. 

\section{Link performance}
\subsection{Time and frequency transfer}

\begin{figure}[htb]
\begin{center}
\resizebox{17cm}{!}{\includegraphics{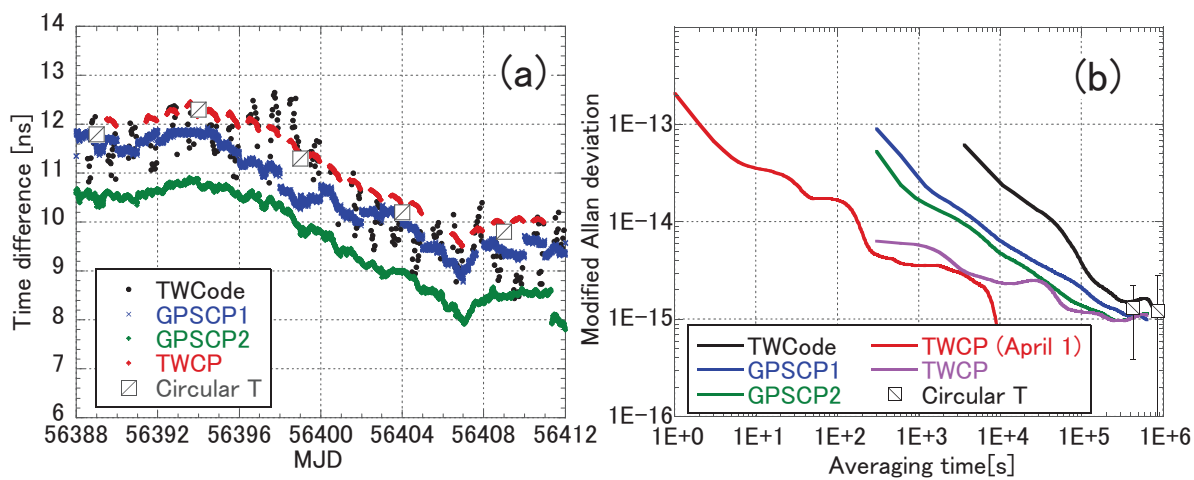}}
\vspace{-177mm}
\caption{(a) Time difference between UTC(NICT) and UTC(PTB) in April 2013,
 (b) Frequency instabilities measured by TWCode, GPSCP and TWCP. The
 frequency instabilities by TWCP in red line and purple line were
 achieved for a short period of one day in April 1 and a long period of
 one month in April, respectively.}
\label{fig:utc_diff}
\end{center}
\end{figure}

\begin{table}[htb]
\caption{GPS receivers used for the NICT-PTB link.}
\label{tbl:gps}
\begin{indented}
\item[]\begin{tabular}{@{}lll}
\br
Caption & Receiver at PTB & Receiver at NICT\\
\mr
GPSCP1 & ASHTECH Z-XII3T & Septentrio PolaRx2 TR\\
GPSCP2 & ASHTECH Z-XII3T & ASHTECH Z-XII2T \\
\br
\end{tabular}
\end{indented}
\end{table}

The TWCP experiment was performed from March to June 2013 
and the time difference between UTC(NICT) and UTC(PTB) was 
measured. 
The result was compared with those achieved by
TWCode and GPSCP for the evaluation. The GPS 
receivers that were used are listed in Table \ref{tbl:gps}, and they performed
dual-frequency code and phase measurements recorded in RINEX 2.1 format. 
There were measurement gaps 
in the TWCP and TWCode results because of the limited 
working time of the satellite
transponder between 12:00 UTC and 22:00 UTC. 
The phase discontinuity in the TWCP was filled 
by an integral multiple
of 
a number 
% the wavelength 
to fit the result of GPSCP2. The numbers of the cycle slips in
the phase information $\phi_{ab}, \phi_{ba}, \phi_{aa}, \phi_{bb}$ are
assumed as $n_{ab}, n_{ba}, n_{aa}, n_{bb}$, which are not necessarily
the same because the phase information are detected by two independent
data-processing sequences at two stations. 
With the amplitude $A=\ 1/(\omega_{+}^{2}-\omega_{-}^{2})$, 
the offset $t_{N}$ in time due to the cycle slips is written
as 
$t_{N}\ =\ A[(n_{ab}-n_{ba})\omega_{+}-(n_{aa}-n_{bb})\omega_{-}] 
=\ A(n\omega_{+}-m\omega_{-})$. Here $n$ and $m$ are arbitrary
integers. Since the minimum resolution of $t_{N}$ is $A\omega_{-}$, 
an integral multiple of $A\omega_{-}$ is used to fill a gap in a TWCP
result. In the subsequent case, it was about 0.8 ps. 

Figure \ref{fig:utc_diff}(a) shows the measurement results obtained
using TWCode,
GPSCP1, GPSCP2, and TWCP. Offset values are applied into the plot for
better visibility. 
The TWCode measurement was performed once every hour. 
In each measurement, the midpoint of the second-order fit for 5 min (300
data points) was calculated. 
The TWCP measurement was continuously performed while the transponder
was available, where data were taken every second. 
The mean values over 5 min are depicted in Figure \ref{fig:utc_diff}. 
The GPSCP results were computed every 5 min as well \cite{Gotoh_gps}. 
The white squares indicate the time difference 
published in the Circular T 304 \cite{circularT}. The time link 
between PTB and NICT adopted for TAI is built on GPSPPP, 
and the same receiver data are used by BIPM as in this study. 
It is clear that the TWCP results agree with that of TWCode 
within the large dispersion of the TWCode data. On the other hand, 
TWCP and GPSCP2 show a good agreement because the phase
ambiguity in the TWCP result was compensated to fit to GPSCP2. Further, 
it appears that it is also almost consistent with GPSCP1. 
A detailed comparison between GPSCP and TWCP is discussed 
in the next subsection. 
Figure \ref{fig:utc_diff}(b) shows their frequency instabilities presented
in the modified Allan deviation, which were from April 1 to April 30,
2013. The TWCP instabilities that were both measured on April 1, 2013 and 
300 s averaged from April 1 to April 30 were depicted. 
We confirmed that an instability of 
$2\times10^{-13}$ at 1 s was possible by TWCP over such a very long 
baseline. The instability obtained by TWCP 
reaches a level below $10^{-14}$ 
after an average time of 200 s. On the other hand, 
those obtained by GPSCP2 and TWCode reached the same level only after
3600 s and 46800 s, respectively. 
At observation times exceeding one day, the comparisons are limited by 
the combined frequency instability of the two time scales/clocks
involved at NICT and PTB. Therefore a detailed analysis using 
double differences is given in the following section.

\subsection{Comparison with GPSCP}

\begin{figure}[htb]
\begin{center}
\resizebox{12cm}{!}{\includegraphics{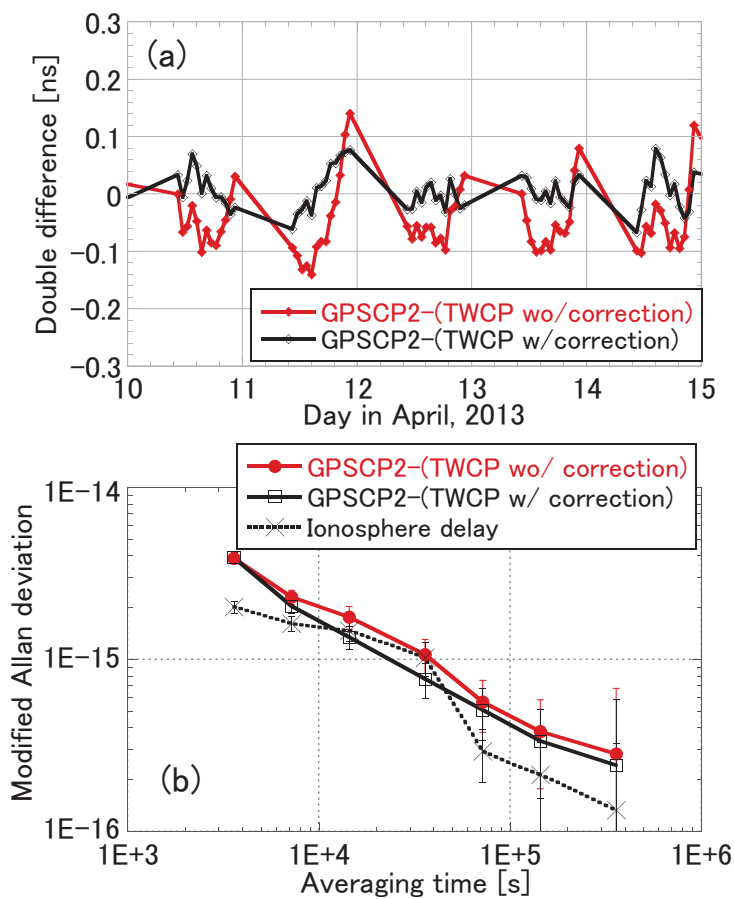}}
\vspace{-54mm}
\caption{(a) 1-h averaged double difference of the NICT-PTB link between
 GPSCP2 and TWCP with and without ionospheric compensation 
 and (b) modified Allan deviation.}
\label{fig:gpscp-twcp_delay}
\end{center}
\end{figure}

\begin{figure}[htb]
\begin{center}
\resizebox{12cm}{!}{\includegraphics{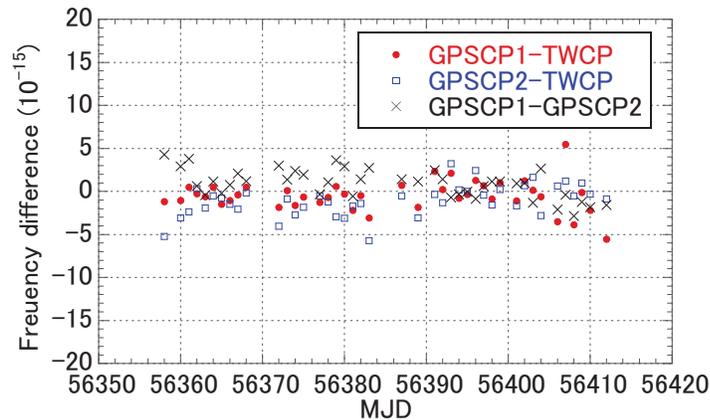}}
\vspace{-116mm}
\caption{Frequency differences for combinations of 
GPSCP1, GPSCP2 and TWCP. They are
 averaged for 1 day (10 hours). Data of GPSCP and TWCP are 
plotted only when both kinds were available. }
\label{fig:freqdiff_gpscp-twcp}
\end{center}
\end{figure}

In order to characterize the NICT-PTB link performance, 
double differences between GPSCP and TWCP solutions are formed which
should be free of clock noise.  
Figure \ref{fig:gpscp-twcp_delay}(a) shows the double differences in
delay between GPSCP2 and TWCP with and without ionospheric compensation. 
The data were averaged for 1 h and then subtracted. 
Here GPSCP1 and GPSCP2 are ionosphere free. The data of GPSCP1 were not
adopted in the plot because there were some discontinuities. 
As shown in Figure \ref{fig:tec_map}(c), the ionospheric delay 
has its maximum close to 0:00 UTC. 
Therefore, the disagreement between 
GPSCP2 and TWCP follows a similar signature. 
However, it was improved by compensating for the ionospheric delay. 
The achieved frequency instabilities expressed 
by the modified Allan deviation are depicted 
in Figure \ref{fig:gpscp-twcp_delay}(b), and the $10^{-16}$
level is reached after 40000 s. 
Those having the ionospheric 
compensation show slightly better values. The dashed line shows the
impact due to the ionospheric delay in the NICT-PTB TWCP link, 
which is at the low $10^{-16}$ level for an averaging time of one day. 
From this result, we deduce that averaging for longer than one day will
be effective for reducing the influence in the case 
that 24-h continuous measurements will be possible. 
\\
\hspace{1em}
The frequency differences for GPSCP1-TWCP, GPSCP2-TWCP, and GPSCP1-GPSCP2 
are shown in Figure \ref{fig:freqdiff_gpscp-twcp} (1-day from 12:00 to
22:00 average values). 
Outliers exceeding three times the standard deviation were previously
removed. 
In the TWCP results, 
we compensated for the ionospheric delay. 
Their differences in March and April 2013 
are listed in Table \ref{tbl:freqdiff}. 
Without the ionospheric compensation, the values of GPSCP1-TWCP and
GPSCP2-TWCP are \(1.24\times10^{-15}\) and \(1.44\times10^{-15}\), 
respectively. 
From these results, it is clear that compensation leads to 
improved consistency. 
The difference between GPSCP1 and GPSCP2 was caused by the GPS receiver used at
NICT. When we limited the data's available time 
from 12:00 to 22:00 in UTC, 
that is, the satellite transponder working time, the frequency difference 
between GPSCP1 and GPSCP2 became larger. %However, for a full day, 
%the difference was smaller. 
From this result, 
the limited measurement period may have contributed to the difference in
frequency. 
%If a 24-h measurement was possible, TWCP and GPSCP may have shown a 
%better agreement. 
In other words, 
any error source that exists with daily variations may remain 
in both TWCP and GPSCP. 
As a result, a frequency offset may be induced by 
limiting the measurement time. 
We used the regional VTEC maps to compensate for the ionospheric delay; 
we have to evaluate their uncertainties more carefully. 
Besides, other error sources should be considered to reduce the 
daily variation. From Table \ref{tbl:freqdiff}, 
we concluded that the systematic uncertainty in frequency comparisons
via the TWCP link was less than $1\times10^{-15}$.

\begin{table}[htb]
\caption{Frequency differences in March and April 2013 ($\times10^{-15}$).}
\label{tbl:freqdiff}
\begin{indented}
\item[]\begin{tabular}{@{}llll}
\br
Caption & Frequency difference & Standard uncertainty & No. of points\\
\mr
GPSCP1-TWCP (12-22h)& -0.56 & 0.33 & 446\\
GPSCP2-TWCP (12-22h)& -0.95 & 0.32 & 453\\
GPSCP1-GPSCP2 (full day) & 0.47 & 0.19 & 1257\\
GPSCP1-GPSCP2 (12-22h) & 0.79 & 0.28 & 442\\
%GPSCP1-TWCP (wo iono)& 1.24 & 0.35 & 446\\
%GPSCP1-TWCP (wo iono)& 1.44 & 0.35 & 453\\
\br
\end{tabular}
\end{indented}
\end{table}

\section{Conclusion}

We performed a TWCP experiment between
NICT and PTB, and we confirmed that TWCP is possible over 
very long baselines, ranging up to about 9,000 km. 
The result showed a good agreement with those measured by TWCode and
GPSCP, and a short-term instability of $2\times10^{-13}$ at 1 s 
was achieved. It is clear that the measurement precision is superior to 
that of TWCode and GPSCP, and it is independent of the baseline length. 
On the other hand, the effect of the ionospheric delay 
is more noticeable over a long baseline. 
We computed the ionospheric delays using the local VTEC maps provided
by ROB and NICT and found that the amplitude reached about 100 ps.
The disagreement between TWCP and GPSCP was decreased by the
compensation. The modified Allan deviation of GPSCP-TWCP was improved 
and reached the $10^{-16}$ level. 
Meanwhile, there remains a frequency disagreement between them at 
the $10^{-16}$ level. One possible explanation is that there 
is an instability due to a daily variation, which induced the frequency
offset caused by the limited measurement time. The long-term 
instability should be further evaluated by performing a 24-h operation. 
In addition, it may be necessary to improve the environment 
in which the instruments are installed, when the high performance of 
TWCP shall be utilized.

\subsection*{Acknowledgments}

We would like to thank H. Maeno and R. Tabuchi at NICT for their
assistance with establishing the link. 
We would also like to thank P. Defraigne at ROB and M. Nishioka at NICT 
for the information regarding the regional VTEC maps.


\section*{References}
\begin{thebibliography}{10}
%\bibitem{book1} Goosens M, Rahtz S and Mittelbach F 1997 {\it The \LaTeX\ Graphics Companion\/} 
%(Reading, MA: Addison-Wesley)
%\bibitem{eps} Reckdahl K 1997 {\it Using Imported Graphics in \LaTeX\ }
%	(search CTAN for the file `epslatex.pdf')

%\bibitem{Kiuchi}
%H. Kiuchi, T. Kawanishi, M. Yamada, T. Sakamoto, M. Tsuchiya, J. Amagai,
%	and M. Izutsu, ``High extinction ratio mach-zehnder modulator
%	applied to a highly stable optical signal generator'', IEEE
%	Trans. Microw. Theory Tech., vol. 55, no. 9, pp. 1964-1972,
%	2007. 
\bibitem{BauchPeik}
E. Peik and A. Bauch, ``More Accurate Clocks - What are They
	Needed for?'', Special Issue / PTB-Mitteilungen, {\bf 119},
	pp. 16-24, 2009. 

\bibitem{Peik}
E. Peik, ``Fundamental constants and units and the search for
	temporal variations'', Nuclear Physics B (Proc. Suppl.),
	203-204, pp. 18-32, 2010.
\bibitem{Frisch}
J. Frisch, D. Bernstein, D. Brown, and E. Cisneros, ``A high stability,
	low noise RF distribution system'', in Proc. Particle
	Accelerator Conf. 2002, vol. 2, 2002, pp. 816-818. 
\bibitem{Poland}
P. Krehlik, L. Sliwczynski, L. Buczek, and M. Lipinski, ``Fiber-Optic
	Joint Time and Frequency Transfer With Active Delay
	Stabilization of the Propagation Delay'', IEEE Trans on
	Instr. Measurement, vol. 61, no. 10, 2844-2851, 2012. 
\bibitem{Rost}
M. Rost, D. Piester, W. Yang, T. Feldmann, T. W\"{u}bbena, A. Bauch, ``Time
	transfer through optical fibers over a distance of 73 km with
	uncertainty below 100 ps'', Metrologia, vol. 49, no. 6,
	pp. 772-778, 2012. 

\bibitem{Lopez}
O. Lopez, A. Kanji, P.-E. Pottie, D. Rovera, J. Achkar, C. Chardonnet,
	A. Amy-Klein, G. Santarelli, ``Simultaneous remote transfer of
	accurate timing and optical frequency over a public fiber
	network'', J. Appl. Phys. B. vol. 110, no. 1, pp. 3-6, 2013. 

\bibitem{WR}
White Rabbit project: http://www.ohwr.org/projects/white-rabbit.\\
P. Moreira, J. Serrano, T. Wolstowski, P. Loschmidt, and G. Gaderer,
	``White rabbit: Sub-nanosecond timing distribution over
	Ethernet'', ISPCS 2009 International Symposium on Precision Clock
	Synchronization for Measurement, Control and Communication,
	pp. 58-62, 2009. 

\bibitem{Yamaguchi}
A. Yamaguchi, M. Fujieda, M. Kumagai, H. Hachisu, S. Nagano, Y. Li,
	T. Ido, T. Takano, M. Takamoto, and H. Katori, ``Direct
	Comparison of Distant Optical Lattice Clocks at the 10$^{-16}$
	Uncertainty'', Applied Physics Express 4, 082203, 2011. 
\bibitem{Matveev}
A. Matveev et al., ``Precision Measurement of the Hydrogen 1S-2S
	Frequency via a 920-km Fiber Link'', \PRL 110, 23081, 2013. 
\bibitem{Science_920km}
K. Predehl, G. Grosche, S. M. F. Raupach, S. Droste, O. Terra, J. Alnis,
	Th. Legero, T. W. Hansch, Th. Udem, R. Holzwarth, H. Schnatz, 
``A 920-Kilometer Optical Fiber Link for Frequency Metrology at the 19th
	Decimal Place'', Science vol. 336, 441-444, 27 APRIL 2012.
\bibitem{Fujieda1}
M. Fujieda, T. Gotoh, F. Nakagawa, R. Tabuchi, M. Aida, and J. Amagai,
	``Carrier-Phase-Based Two-Way Satellite Time and Frequency
	Transfer'', IEEE Trans. Ultrason. Ferroelectr. Freq. Control,
	vol. 59, no. 12, 2625-2630, 2012. 
\bibitem{PiesterSchnatz}
D. Piester, H. Schnatz, ``Novel techniques for remote time and frequency
	comparisons'', PTB-Mitteilungen, vol. 119, pp. 33-44 (Special
	issue), 2009. 
\bibitem{BIPM}
G. Panfilo and E. F. Arias, ``Algorithms for International Atomic
	Time'', IEEE Trans. Ultrason. Ferroelectr. Freq. Control,
	vol. 57, no. 1, pp. 140-150, 2010.
\bibitem{Kirchner}
D. Kirchner, ``Two-way satellite time and frequency transfer (TWSTFT):
	principle, implementation, and current performance,'' {\sl Rev. of
	Radio Science}, Oxford University Press, pp. 27-44, 1999. 
\bibitem{GPS_geodesy}
A. Kleusberg, ``Atmospheris Models from GPS'', section 15, {\it GPS
	for geodesy}, 2nd edition, Berlin: Springer, 1998. 

\bibitem{Piester2008}
D. Piester, A. Bauch, J. Becker, E. Staliuniene, and C. Schlunegger,
	``On Measurement Noise in the European TWSTFT Network'', IEEE
	Trans. Ultrason. Ferroelectr. Freq. Control, vol. 55, no. 9,
	pp. 1906-1912, 2008.
\bibitem{Schaefer}
W. Sch\"{a}fer, A. Pawlitzki, and T. Kuhn, ``New Trends in Two-Way
	Time and Frequency Transfer via Satellite'', in Proc. 31$^{th}$
	Annual Precise Time and Time Interval (PTTI) Meeting,
	pp. 505-514, 1999.

\bibitem{Fonville}
B. Fonville, D. Matsakis, A. Pawlitzki, W. Sch\"{a}efer, ``Development of 
	Carrier-phase-based Two-way Satellite Time and Frequency 
	Transfer (TWSTFT)'', in Proc. 36$^{th}$ Annual Precise Time and
	Time Interval (PTTI) Meeting, pp. 149-164, 2004. 

\bibitem{Tom}
T. E. Parker and V. Zhang, ``Sources of Instabilities in Two-Way
	Satellite Time Transfer'', in Proc. IEEE Int. Freq. Control
	Sympo. and 37$^{th}$ Annual Precise Time and Interval (PTTI) Meeting,
	pp. 745-751, 2005. 

\bibitem{Amagai}
J. Amagai and T. Gotoh, ``Development of Two-Way Time and Frequency
	Transfer System with Dual Pseudo Random Noises'', Journal of the
	National Institute of Information and Communications Technology,
	Vol. 57, Nos. 3/4, pp.197-207, 2010. 
\bibitem{Gotoh}
T. Gotoh, J. Amagai, T. Hobiger, M. Fujieda, and M. Aida, ``Development
	of a GPU-Based Two-Way Time Transfer Modem'', IEEE
	Trans. Instr. Meas., vol. 60, no. 7, pp. 2495-2499, 2011.

\bibitem{VLBI}
T. Kondo, Y. Koyama, R. Ichikawa, M. Sekido, E. Kawai, M. Kimura,
	``Development of the K5/VSSP system'', J. Geodetic Soc. Japan,
	vol. 54, no.4, pp. 233-248, 2008. 

\bibitem{Bauch2}
A. Bauch, S. Weyers, D. Piester, E. Staliuniene and W. Yang,
	``Generation of UTC(PTB) as a fountain-clock based time scale'',
	Metrologia, vol. 49, pp. 180-188, 2012. 
\bibitem{Nakagawa}
F. Nakagawa, Y. Hanado, H. Ito, N. Kotake, M. Kumagai, K. Imamura, and
	Y. Koyama, ``Summary and Improvement of Japan Standard Time
	Generation System'', Journal of the
	National Institute of Information and Communications Technology,
	vol. 57, Nos. 3/4, pp. 17-27, 2010. 

%M. Kimura, ``The implementation of the PC based giga bit VLBI system'',
%	IVS Technical Development Center News, no. 21, pp. 31-33, 2002. 


\bibitem{Bauch1}
A. Bauch, D. Piester, M. Fujieda, W. Lewandowski, ``Directive for
	operational use and data handling in two-way satellite time and
	frequency transfer (TWSTFT)'', Rapport BIPM-2011/01, 2011.
\bibitem{CODE}
http://www.aiub.unibe.ch/content/research/satellite\_geodesy/code\_research/index\_eng.htm

\bibitem{ionex}
S. Schaer and W. Gurtner, ``IONEX: The IONosphere Map EXchange Format
	Version 1'', Proc. of the IGS AC Workshop, Darmstadt, Germany,
	February 9-11, 1998.
\bibitem{ROB}
http://www.gnss.be/Atmospheric\_Maps/ionospheric\_maps.php \\
N. Bergeot, J.-M. Chevalier, L. Benoit, C. Bruyninx, J. Legrand,
	E. Pottiaux, Q. Baire, P. Defraigne, ``Near Real Time Ionosphere
	Models from European Permanent Network GPS Data'', EUREF Annual
	Sym, 24-27/05/2011, Chisinau, Moldova. 
\bibitem{NICT_TEC}
http://wdc.nict.go.jp/IONO/gps-tec/tecv/ \\
G. Ma and T. Maruyama, ``Derivation of TEC and estimation of
	instrumental biases from GEONET in Japan'', Ann. Geophys.,
	pp. 2083-2093, 2003. 
\bibitem{EUREF}
http://epncb.oma.be

\bibitem{GEONET}
S. Miyazaki, T. Saito, M. Sasaki, Y. Hatanaka, and Y. Iimura,
	``Expansion of GSI's nationwide GPS array'',
	Bull. Geogr. Surv. Inst., no. 43, pp. 23-34, 1997.\\ 
Geographical Survey Institute, Japan, ``GEONET (GPS Earth Observation
	Network System) and its Prospect'', Journal of the Geodetic
	Society of Japan, Vol. 50, No. 2, pp. 53-65, 2004. (in Japanese)
\bibitem{CGGTTS}
D.W. Allan and C. Thomas, ``Technical Directives for Standardization of
	GPS Time Receiver Software'', Metrologia, vol. 31, pp. 67-79,
	1994. 

\bibitem{Piester1}
D. Piester, A. Bauch, M. Fujieda, T. Gotoh, M. Aida,
	H. Maeno,M. Hosokawa and S. H. Yang, ``Studies on Instabilities 
	in Long-baseline Two-way Satellite Time and Frequency Transfer 
	(TWSTFT) Including a Troposphere Delay Model'', Proc. of
	39$^{th}$ Annual Precise Time and Interval (PTTI) Meeting,
	pp. 211-222, 2007. 
\bibitem{NORAD}
http://celestrak.com/NORAD/documentation/tle-fmt.asp
\bibitem{Gotoh_gps}
T. Gotoh, M. Fujieda, J. Amagai, ``Comparison study of GPS carrier phase 
and two-way time and frequency transfer'', Proc. of Joint EFTF and IFCS,
	2007. 
\bibitem{circularT}
ftp://ftp2.bipm.org/pub/tai/publication/cirt/cirt.304 

\end{thebibliography}
\end{document}